\documentclass[letterpaper,compsoc,twoside]{IEEEtran}
\usepackage{fixltx2e} 
\usepackage{cmap} 
\usepackage{ifthen}
\usepackage[T1]{fontenc}
\usepackage[utf8]{inputenc}
\usepackage{amsmath}

\usepackage[font={small,it},labelfont=bf]{caption}
\usepackage{float}

\pdfoutput=1
\usepackage{scipy}
\makeatletter
\def\PY@reset{\let\PY@it=\relax \let\PY@bf=\relax%
    \let\PY@ul=\relax \let\PY@tc=\relax%
    \let\PY@bc=\relax \let\PY@ff=\relax}
\def\PY@tok#1{\csname PY@tok@#1\endcsname}
\def\PY@toks#1+{\ifx\relax#1\empty\else%
    \PY@tok{#1}\expandafter\PY@toks\fi}
\def\PY@do#1{\PY@bc{\PY@tc{\PY@ul{%
    \PY@it{\PY@bf{\PY@ff{#1}}}}}}}
\def\PY#1#2{\PY@reset\PY@toks#1+\relax+\PY@do{#2}}

\expandafter\def\csname PY@tok@gd\endcsname{\def\PY@tc##1{\textcolor[rgb]{0.63,0.00,0.00}{##1}}}
\expandafter\def\csname PY@tok@gu\endcsname{\let\PY@bf=\textbf\def\PY@tc##1{\textcolor[rgb]{0.50,0.00,0.50}{##1}}}
\expandafter\def\csname PY@tok@gt\endcsname{\def\PY@tc##1{\textcolor[rgb]{0.00,0.27,0.87}{##1}}}
\expandafter\def\csname PY@tok@gs\endcsname{\let\PY@bf=\textbf}
\expandafter\def\csname PY@tok@gr\endcsname{\def\PY@tc##1{\textcolor[rgb]{1.00,0.00,0.00}{##1}}}
\expandafter\def\csname PY@tok@cm\endcsname{\let\PY@it=\textit\def\PY@tc##1{\textcolor[rgb]{0.25,0.50,0.56}{##1}}}
\expandafter\def\csname PY@tok@vg\endcsname{\def\PY@tc##1{\textcolor[rgb]{0.73,0.38,0.84}{##1}}}
\expandafter\def\csname PY@tok@m\endcsname{\def\PY@tc##1{\textcolor[rgb]{0.13,0.50,0.31}{##1}}}
\expandafter\def\csname PY@tok@mh\endcsname{\def\PY@tc##1{\textcolor[rgb]{0.13,0.50,0.31}{##1}}}
\expandafter\def\csname PY@tok@cs\endcsname{\def\PY@tc##1{\textcolor[rgb]{0.25,0.50,0.56}{##1}}\def\PY@bc##1{\setlength{\fboxsep}{0pt}\colorbox[rgb]{1.00,0.94,0.94}{\strut ##1}}}
\expandafter\def\csname PY@tok@ge\endcsname{\let\PY@it=\textit}
\expandafter\def\csname PY@tok@vc\endcsname{\def\PY@tc##1{\textcolor[rgb]{0.73,0.38,0.84}{##1}}}
\expandafter\def\csname PY@tok@il\endcsname{\def\PY@tc##1{\textcolor[rgb]{0.13,0.50,0.31}{##1}}}
\expandafter\def\csname PY@tok@go\endcsname{\def\PY@tc##1{\textcolor[rgb]{0.20,0.20,0.20}{##1}}}
\expandafter\def\csname PY@tok@cp\endcsname{\def\PY@tc##1{\textcolor[rgb]{0.00,0.44,0.13}{##1}}}
\expandafter\def\csname PY@tok@gi\endcsname{\def\PY@tc##1{\textcolor[rgb]{0.00,0.63,0.00}{##1}}}
\expandafter\def\csname PY@tok@gh\endcsname{\let\PY@bf=\textbf\def\PY@tc##1{\textcolor[rgb]{0.00,0.00,0.50}{##1}}}
\expandafter\def\csname PY@tok@ni\endcsname{\let\PY@bf=\textbf\def\PY@tc##1{\textcolor[rgb]{0.84,0.33,0.22}{##1}}}
\expandafter\def\csname PY@tok@nl\endcsname{\let\PY@bf=\textbf\def\PY@tc##1{\textcolor[rgb]{0.00,0.13,0.44}{##1}}}
\expandafter\def\csname PY@tok@nn\endcsname{\let\PY@bf=\textbf\def\PY@tc##1{\textcolor[rgb]{0.05,0.52,0.71}{##1}}}
\expandafter\def\csname PY@tok@no\endcsname{\def\PY@tc##1{\textcolor[rgb]{0.38,0.68,0.84}{##1}}}
\expandafter\def\csname PY@tok@na\endcsname{\def\PY@tc##1{\textcolor[rgb]{0.25,0.44,0.63}{##1}}}
\expandafter\def\csname PY@tok@nb\endcsname{\def\PY@tc##1{\textcolor[rgb]{0.00,0.44,0.13}{##1}}}
\expandafter\def\csname PY@tok@nc\endcsname{\let\PY@bf=\textbf\def\PY@tc##1{\textcolor[rgb]{0.05,0.52,0.71}{##1}}}
\expandafter\def\csname PY@tok@nd\endcsname{\let\PY@bf=\textbf\def\PY@tc##1{\textcolor[rgb]{0.33,0.33,0.33}{##1}}}
\expandafter\def\csname PY@tok@ne\endcsname{\def\PY@tc##1{\textcolor[rgb]{0.00,0.44,0.13}{##1}}}
\expandafter\def\csname PY@tok@nf\endcsname{\def\PY@tc##1{\textcolor[rgb]{0.02,0.16,0.49}{##1}}}
\expandafter\def\csname PY@tok@si\endcsname{\let\PY@it=\textit\def\PY@tc##1{\textcolor[rgb]{0.44,0.63,0.82}{##1}}}
\expandafter\def\csname PY@tok@s2\endcsname{\def\PY@tc##1{\textcolor[rgb]{0.25,0.44,0.63}{##1}}}
\expandafter\def\csname PY@tok@vi\endcsname{\def\PY@tc##1{\textcolor[rgb]{0.73,0.38,0.84}{##1}}}
\expandafter\def\csname PY@tok@nt\endcsname{\let\PY@bf=\textbf\def\PY@tc##1{\textcolor[rgb]{0.02,0.16,0.45}{##1}}}
\expandafter\def\csname PY@tok@nv\endcsname{\def\PY@tc##1{\textcolor[rgb]{0.73,0.38,0.84}{##1}}}
\expandafter\def\csname PY@tok@s1\endcsname{\def\PY@tc##1{\textcolor[rgb]{0.25,0.44,0.63}{##1}}}
\expandafter\def\csname PY@tok@gp\endcsname{\let\PY@bf=\textbf\def\PY@tc##1{\textcolor[rgb]{0.78,0.36,0.04}{##1}}}
\expandafter\def\csname PY@tok@sh\endcsname{\def\PY@tc##1{\textcolor[rgb]{0.25,0.44,0.63}{##1}}}
\expandafter\def\csname PY@tok@ow\endcsname{\let\PY@bf=\textbf\def\PY@tc##1{\textcolor[rgb]{0.00,0.44,0.13}{##1}}}
\expandafter\def\csname PY@tok@sx\endcsname{\def\PY@tc##1{\textcolor[rgb]{0.78,0.36,0.04}{##1}}}
\expandafter\def\csname PY@tok@bp\endcsname{\def\PY@tc##1{\textcolor[rgb]{0.00,0.44,0.13}{##1}}}
\expandafter\def\csname PY@tok@c1\endcsname{\let\PY@it=\textit\def\PY@tc##1{\textcolor[rgb]{0.25,0.50,0.56}{##1}}}
\expandafter\def\csname PY@tok@kc\endcsname{\let\PY@bf=\textbf\def\PY@tc##1{\textcolor[rgb]{0.00,0.44,0.13}{##1}}}
\expandafter\def\csname PY@tok@c\endcsname{\let\PY@it=\textit\def\PY@tc##1{\textcolor[rgb]{0.25,0.50,0.56}{##1}}}
\expandafter\def\csname PY@tok@mf\endcsname{\def\PY@tc##1{\textcolor[rgb]{0.13,0.50,0.31}{##1}}}
\expandafter\def\csname PY@tok@err\endcsname{\def\PY@bc##1{\setlength{\fboxsep}{0pt}\fcolorbox[rgb]{1.00,0.00,0.00}{1,1,1}{\strut ##1}}}
\expandafter\def\csname PY@tok@kd\endcsname{\let\PY@bf=\textbf\def\PY@tc##1{\textcolor[rgb]{0.00,0.44,0.13}{##1}}}
\expandafter\def\csname PY@tok@ss\endcsname{\def\PY@tc##1{\textcolor[rgb]{0.32,0.47,0.09}{##1}}}
\expandafter\def\csname PY@tok@sr\endcsname{\def\PY@tc##1{\textcolor[rgb]{0.14,0.33,0.53}{##1}}}
\expandafter\def\csname PY@tok@mo\endcsname{\def\PY@tc##1{\textcolor[rgb]{0.13,0.50,0.31}{##1}}}
\expandafter\def\csname PY@tok@mi\endcsname{\def\PY@tc##1{\textcolor[rgb]{0.13,0.50,0.31}{##1}}}
\expandafter\def\csname PY@tok@kn\endcsname{\let\PY@bf=\textbf\def\PY@tc##1{\textcolor[rgb]{0.00,0.44,0.13}{##1}}}
\expandafter\def\csname PY@tok@o\endcsname{\def\PY@tc##1{\textcolor[rgb]{0.40,0.40,0.40}{##1}}}
\expandafter\def\csname PY@tok@kr\endcsname{\let\PY@bf=\textbf\def\PY@tc##1{\textcolor[rgb]{0.00,0.44,0.13}{##1}}}
\expandafter\def\csname PY@tok@s\endcsname{\def\PY@tc##1{\textcolor[rgb]{0.25,0.44,0.63}{##1}}}
\expandafter\def\csname PY@tok@kp\endcsname{\def\PY@tc##1{\textcolor[rgb]{0.00,0.44,0.13}{##1}}}
\expandafter\def\csname PY@tok@w\endcsname{\def\PY@tc##1{\textcolor[rgb]{0.73,0.73,0.73}{##1}}}
\expandafter\def\csname PY@tok@kt\endcsname{\def\PY@tc##1{\textcolor[rgb]{0.56,0.13,0.00}{##1}}}
\expandafter\def\csname PY@tok@sc\endcsname{\def\PY@tc##1{\textcolor[rgb]{0.25,0.44,0.63}{##1}}}
\expandafter\def\csname PY@tok@sb\endcsname{\def\PY@tc##1{\textcolor[rgb]{0.25,0.44,0.63}{##1}}}
\expandafter\def\csname PY@tok@k\endcsname{\let\PY@bf=\textbf\def\PY@tc##1{\textcolor[rgb]{0.00,0.44,0.13}{##1}}}
\expandafter\def\csname PY@tok@se\endcsname{\let\PY@bf=\textbf\def\PY@tc##1{\textcolor[rgb]{0.25,0.44,0.63}{##1}}}
\expandafter\def\csname PY@tok@sd\endcsname{\let\PY@it=\textit\def\PY@tc##1{\textcolor[rgb]{0.25,0.44,0.63}{##1}}}

\makeatother

\providecommand*{\DUrole}[2]{%
  \ifcsname DUrole#1\endcsname%
    \csname DUrole#1\endcsname{#2}%
  \else
    \ifcsname docutilsrole#1\endcsname%
      \csname docutilsrole#1\endcsname{#2}%
    \else%
      #2%
    \fi%
  \fi%
}

\ifthenelse{\isundefined{\DUlegend}}{
  \newenvironment{DUlegend}{\small}{}
}{}

\providecommand*{\DUroletitlereference}[1]{\textsl{#1}}

\ifthenelse{\isundefined{\hypersetup}}{
  \usepackage[colorlinks=true,linkcolor=blue,urlcolor=blue]{hyperref}
  \urlstyle{same} 
}{}

\begin{document}
\newcounter{footnotecounter}\title{Wyrm, A Pythonic Toolbox for Brain-Computer Interfacing}\author{Bastian Venthur$^{\setcounter{footnotecounter}{1}\fnsymbol{footnotecounter}\setcounter{footnotecounter}{2}\fnsymbol{footnotecounter}}$%
          \setcounter{footnotecounter}{1}\thanks{\fnsymbol{footnotecounter} %
          Corresponding author: \protect\href{mailto:bastian.venthur@tu-berlin.de}{bastian.venthur@tu-berlin.de}}\setcounter{footnotecounter}{2}\thanks{\fnsymbol{footnotecounter} Berlin Institute of Technology}, Benjamin Blankertz$^{\setcounter{footnotecounter}{2}\fnsymbol{footnotecounter}}$\thanks{%

          \noindent%
          Copyright\,\copyright\,2014 Bastian Venthur et al. This is an open-access article distributed under the terms of the Creative Commons Attribution License, which permits unrestricted use, distribution, and reproduction in any medium, provided the original author and source are credited. http://creativecommons.org/licenses/by/3.0/%
        }}\maketitle
          \renewcommand{\leftmark}{PROC. OF THE 7th EUR. CONF. ON PYTHON IN SCIENCE (EUROSCIPY 2014)}
          \renewcommand{\rightmark}{WYRM, A PYTHONIC TOOLBOX FOR BRAIN-COMPUTER INTERFACING}

\setcounter{page}{23}
\newcommand*{\docutilsroleref}{\ref}
\newcommand*{\docutilsrolelabel}{\label}
\AtEndDocument{\cleardoublepage}
\begin{abstract}A Brain-Computer Interface (BCI) is a system that measures central nervous
system activity and translates the recorded data into an output suitable for
a computer to use as an input signal. Such a BCI system consists of three
parts, the signal acquisition, the signal processing and the
feedback/stimulus presentation.

In this paper we present Wyrm, a signal processing toolbox for BCI in
Python. Wyrm is applicable to a broad range of neuroscientific problems and
capable for running online experiments in real time and off-line data
analysis and visualisation.\end{abstract}

\section{Introduction%
  \label{introduction}%
}

In the last years Python has gained more and more traction in the scientific
community. Projects like Numpy \cite{Numpy}, SciPy \cite{SciPy}, and Matplotlib
\cite{Matplotlib} have created a strong foundation for scientific computing in
Python and machine learning packages like Scikit-learn \cite{Scikit-learn} or
packages for data analysis like Pandas \cite{Pandas} are building on top of it.
Although in recent years Python toolboxes like SCoT for EEG source connectivity
\cite{Billinger}, or MNE-Python for MEG and EEG data analysis \cite{Gramfort} were
published, Matlab seems still to be the dominant programming language in the
brain-computer interface (BCI) community.

A BCI is a system that measures central nervous system activity and translates
the recorded data into an output suitable for a computer to use as an input
signal. Or slightly less abstract: A BCI is a communication channel that allows
for direct control of a computer by the power of thoughts.

A BCI system consists of three parts (Figure \DUrole{ref}{bcisystem}): a \emph{signal
acquisition} part that is connected to the measuring hardware (e.g. EEG) and
provides the raw data to the rest of the BCI system. The \emph{signal processing}
part receives the data from the signal acquisition and translates the data into
the intent. The feedback/stimulus presentation part translates the intent into
an action.\begin{figure}[]\noindent\makebox[\columnwidth][c]{\includegraphics[width=\columnwidth]{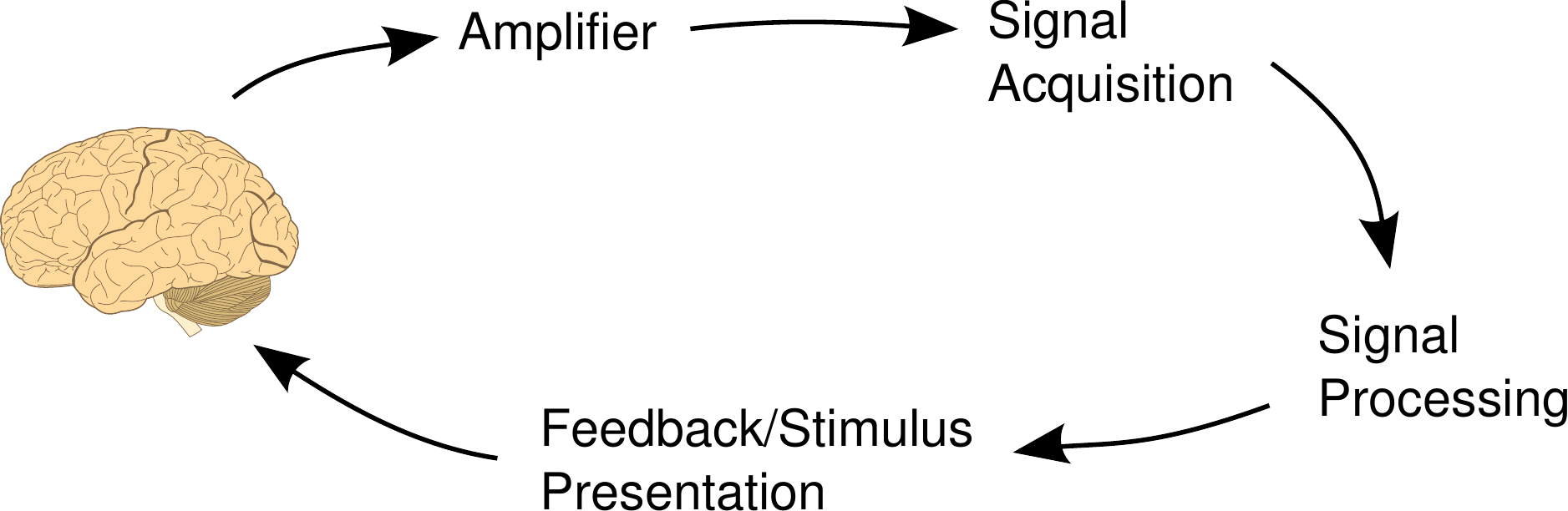}}
\caption{Overview of an online BCI system.
\DUrole{label}{bcisystem}}
\end{figure}

In this paper we present Wyrm, an open source BCI toolbox in Python. Wyrm
corresponds to the signal processing part of the aforementioned BCI system. Wyrm
is applicable to a wide range of neuroscientific problems. It can be used as a
toolbox for analysis and visualization of neurophysiological data (e.g. EEG,
ECoG, fMRI, or NIRS) and it is suitable for real-time online experiments. In
Wyrm we implemented dozens of methods, covering a broad range of aspects for
off-line analysis and online experiments.

In the following sections we will explain Wyrm's fundamental data structure, its
design principles and give an overview of the available methods of the toolbox.
We'll show you where to find the documentation as well as some extensive
examples.

\section{Design%
  \label{design}%
}

All methods in the toolbox revolve around a data structure that is used
throughout the toolbox to store various kinds of data. The data structure dubbed
\DUroletitlereference{Data}, is an object containing an n-dimensional numpy array that represents the
actual data to be stored and some meta data. The meta data describes for each
dimension of the numpy array, the name of the dimension, the names of the single
columns and the unit of the data.

Let's assume we have a record of previously recorded EEG data. The data was
recorded with 20 channels and consists of 30 samples. The data itself can be
represented as a 2-dimensional array with the shape (30, 20). The names for the
dimensions are 'time' and 'channels', the units are 'ms' and '\#' (we use '\#' as
a pseudo unit for enumeration of things), and the description of the columns
would be two arrays: one array of length 30 containing the time stamps for each
sample and another array of length 20 containing the channel names. This data
structure can hold all kinds of BCI related data: continuous data, epoched data,
spectrograms, feature vectors, etc.

We purposely kept the meta data at a minimum, as each operation that modifies
the data also has to check if the meta data is still consistent. While it might
be desirable to have more meta data, this would also lead to more housekeeping
code which makes the code less readable and more error prone. The data
structure, however, can be extended as needed by adding new attributes
dynamically at runtime. All toolbox methods are written in a way that they
ignore unknown attributes and never throw them away.

We tried very hard to to keep the syntax and semantics of the toolbox methods
consistent. Each method obeys a small set of rules: (1) Methods never modify
their input arguments. This allows for a functional style of programming which
is in our opinion well suited when diving into the data. (2) A Method never
modifies attributes of \DUroletitlereference{Data} objects which are unrelated to the functioning of
that method. Especially does it never remove additional or unknown attributes.
(3) If a method operates on a specific axis of a \DUroletitlereference{Data} object, it by default
obeys a convention about the default ordering of the axis but allows for
changing the index of the axis by means of a default arguments.

\section{Toolbox Methods%
  \label{toolbox-methods}%
}

The toolbox contains a few data structures (\DUroletitlereference{Data}, \DUroletitlereference{RingBuffer} and
\DUroletitlereference{BlockBuffer}), I/O-methods for loading and storing data in foreign formats and
off course the actual toolbox algorithms. The list of algorithms includes:
channel selection, IIR filters, sub-sampling, spectrograms, spectra, baseline
removal for signal processing; Common Spatial Patterns (CSP) {[}Ramoser{]}, Source
Power Co-modulation (SPoC) {[}Dähne{]}, classwise average, jumping means, signed
$r^2$-values for feature extraction; Linear Discriminant Analyis (LDA)
with and without shrinkage for machine learning \cite{Blankertz}, and many more. It
is worth mentioning that with scikit-learn you have a wide range of machine
learning algorithms readily at your disposal. Our data format is very compatible
with scikit-learn and one can usually apply the algorithms without any data
conversion step at all.

The toolbox also includes plotting facilities that make it easy to quickly
generate useful plots out of neurophysiological data. Those methods include
scalp plots (Figure \DUrole{ref}{scalp}), time courses (Figure \DUrole{ref}{average}), signed
$r^2$ plots (Figure \DUrole{ref}{r2}), and more.\begin{figure}[]\noindent\makebox[\columnwidth][c]{\includegraphics[width=\columnwidth]{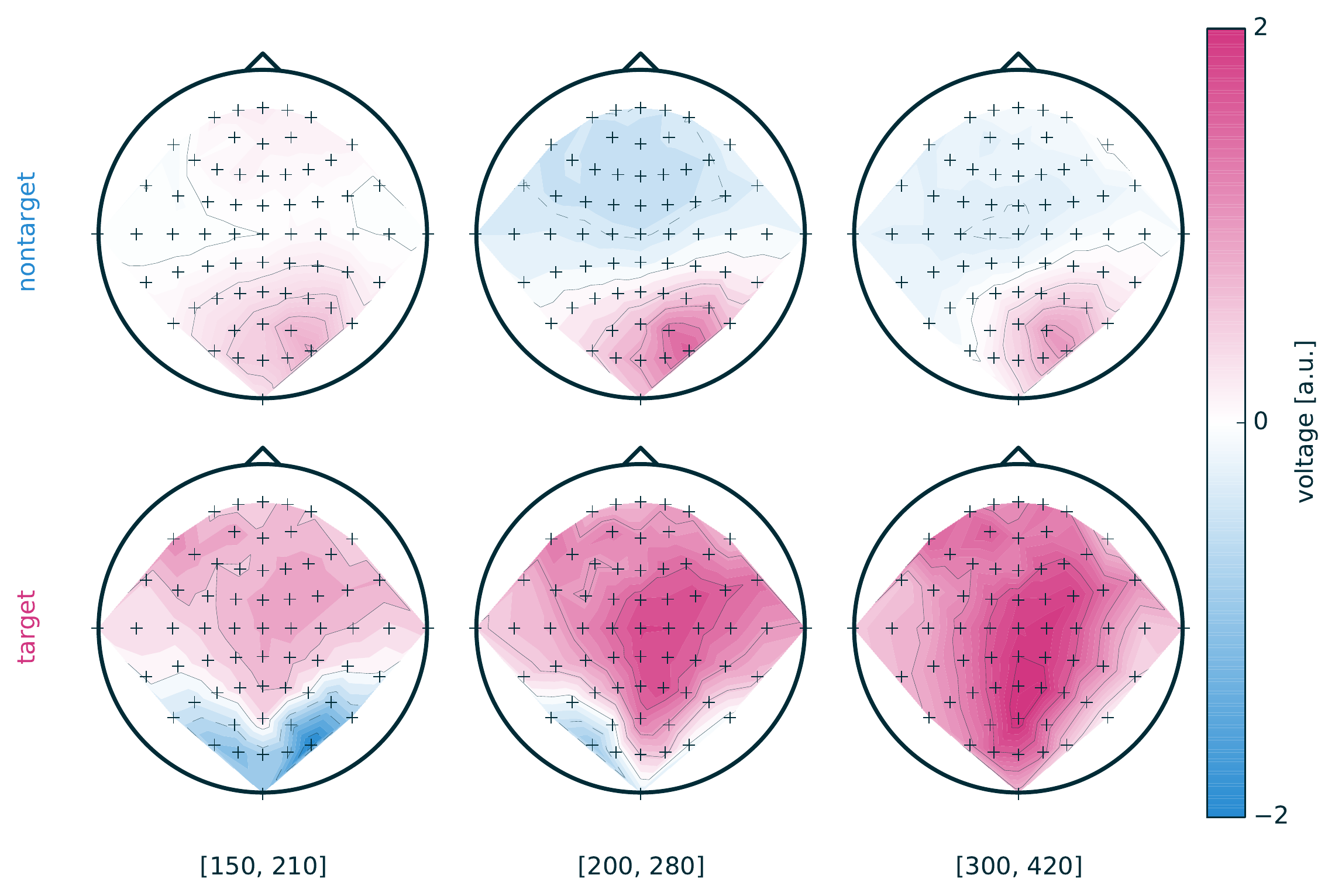}}
\caption{Example scalp plots. The plots show the spatial topography of the average
voltage distribution for different time intervals and channels.}
\begin{DUlegend}

\DUrole{label}{scalp}\end{DUlegend}
\end{figure}\begin{figure}[]\noindent\makebox[\columnwidth][c]{\includegraphics[width=\columnwidth]{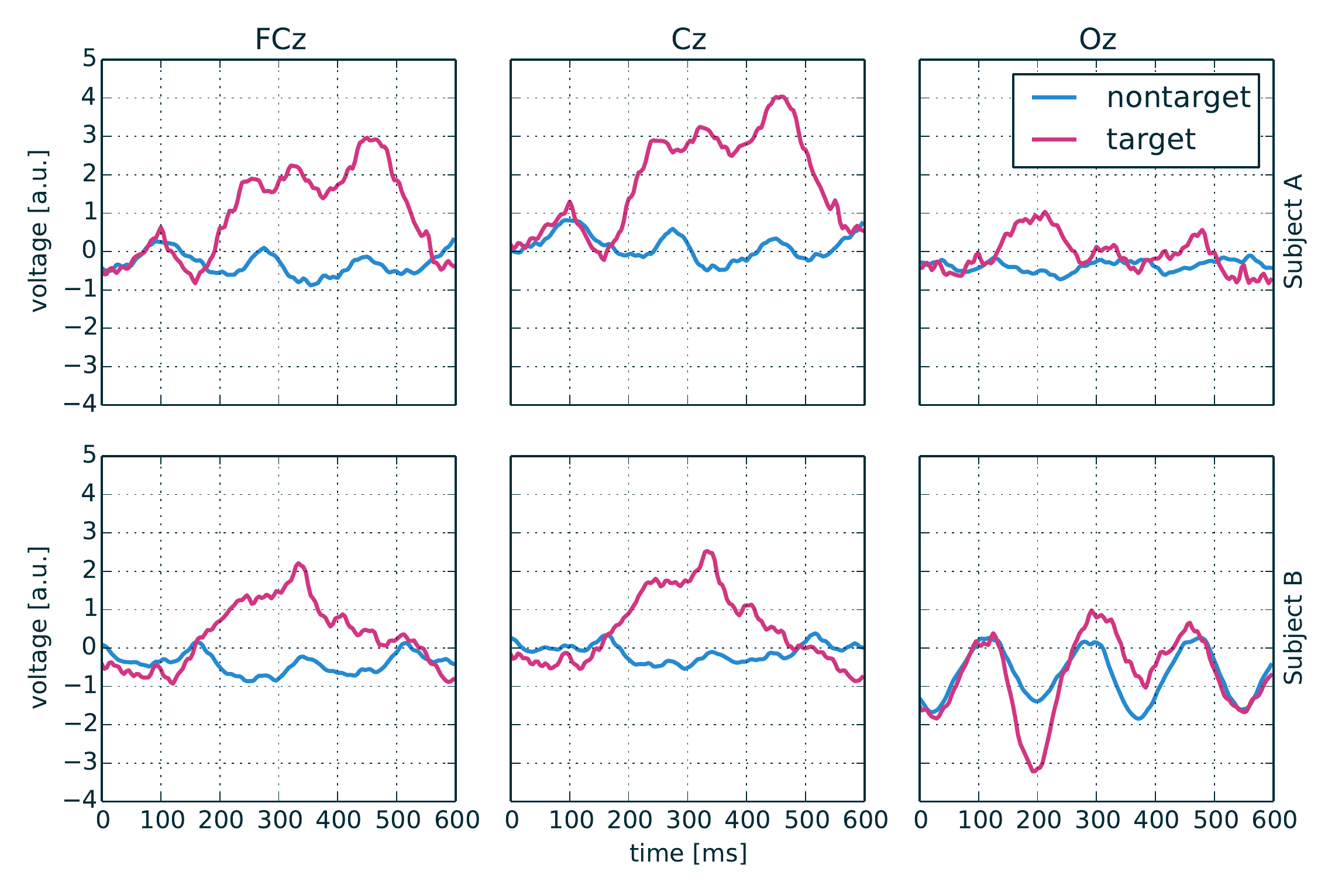}}
\caption{Example time course plots for three selected channels.}
\begin{DUlegend}

\DUrole{label}{average}\end{DUlegend}
\end{figure}\begin{figure}[]\noindent\makebox[\columnwidth][c]{\includegraphics[width=\columnwidth]{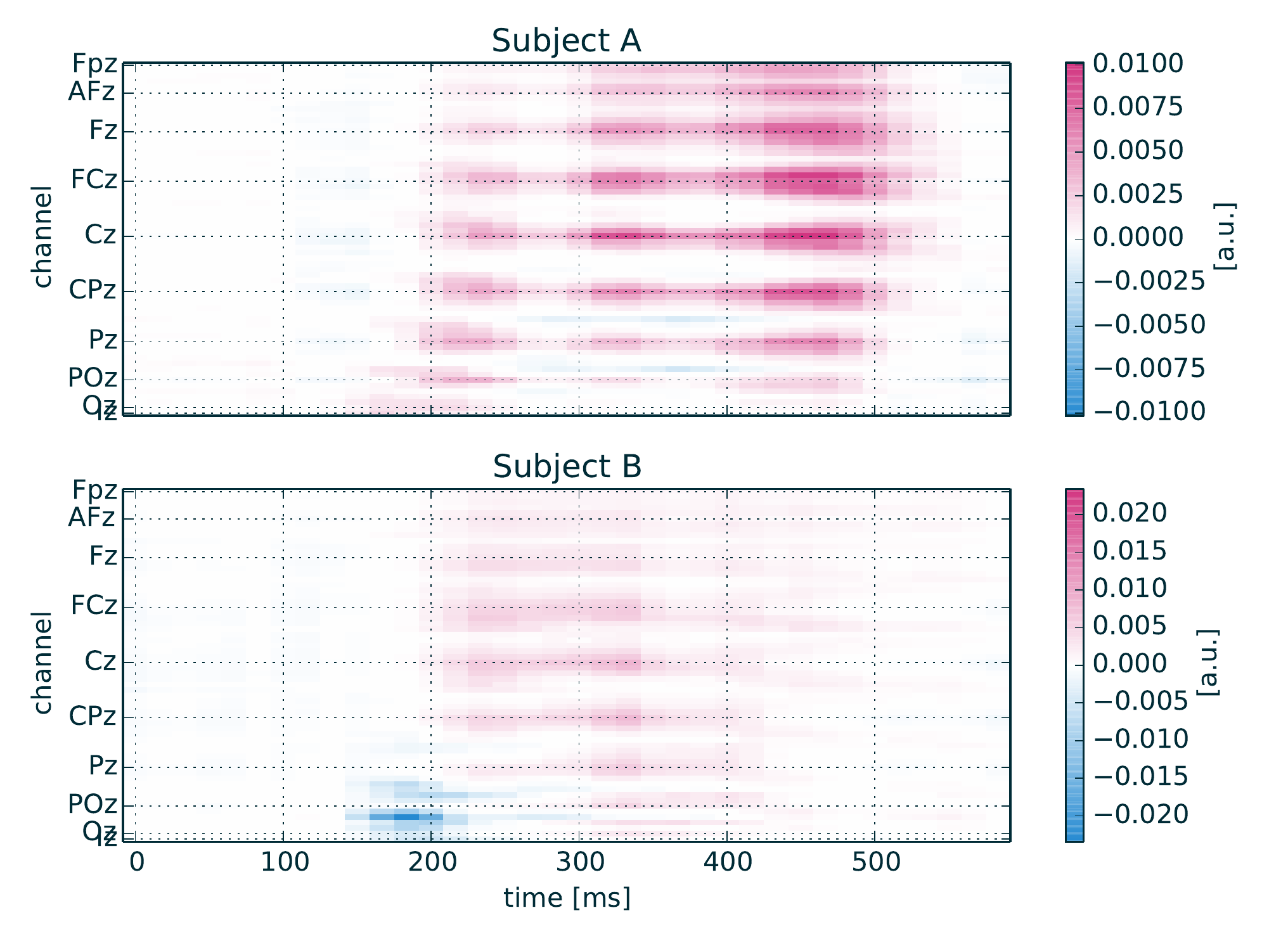}}
\caption{Example signed $r^2$-plots. The channels are sorted from frontal to
occipital and within each row from left to right.}
\begin{DUlegend}

\DUrole{label}{r2}\end{DUlegend}
\end{figure}

\section{Unit Testing%
  \label{unit-testing}%
}

Since the correctness of its methods is crucial for a toolbox, we used unit
testing to ensure all methods work as intended. In our toolbox \emph{each} method is
tested respectively by at least a handful of test cases to ensure that the
methods calculate the correct results, throw the expected errors if necessary,
etc. The total amount of code for all tests is roughly 2-3 times bigger than the
amount code for the toolbox methods.

\section{Documentation%
  \label{documentation}%
}

As a software toolbox would be hard to use without proper documentation, we
provide documentation that consists of readable prose and extensive API
documentation (\url{http://venthur.github.io/wyrm/}). Each method of the toolbox is
thoroughly documented and has usually a short summary, a detailed description of
the algorithm, a list of expected inputs, return values and exceptions, as well
as cross references to related methods in- or outside the toolbox and example
code to demonstrate how to use the method.

\section{Examples%
  \label{examples}%
}

To show how to use the toolbox realistic scenarios we provide two off-line
analysis scripts, where we demonstrate how to use the toolbox to complete two
tasks from the BCI Competition III \cite{BCIComp3}. The first example uses
Electrocorticography (ECoG) recordings provided by the
Eberhard-Karls-Universität Tübingen. The time series where picked up by a 8x8
ECoG platinum grid which was placed on the contralateral, right motor cortex.

During the experiment the subject had to perform imagined movements of either
the left small finger or the tongue. Each trial consisted of either an imagined
finger- or tongue movement and was recorded for a duration of 3 seconds. The
recordings in the data set start at 0.5 seconds after the visual cue had ended
to avoid visual evoked potentials, being reflected by the data. It is worth
noting that the training- and test data were recorded on the same subject but
with roughly one week between both recordings. The data set consists of 278
trials of training data and 100 trials of test data. During the BCI Competition
only the labels (finger or tongue movement) for the training data were
available. The task for the competition was to use the training data and its
labels to predict the 100 labels of the test data. Since the competition is
over, we also had the true labels for the test data, so we could calculate and
compare the accuracy of our results. For this experiment our classification
accuracy was 92\% which is comparable with the winners of the competition whose
accuracy was: 91\%, 87\%, and 86\%.

The second data set uses Electroencephalography (EEG) recordings, provided by
the Wadsworth Center, NYS Department of Health, USA. The data were acquired
using BCI2000’s Matrix Speller paradigm, originally described in \cite{Donchin}. The
subject had to focus on one out of 36 different characters, arranged in a 6x6
matrix. The rows and columns were successively and randomly intensified. Two out
of 12 intensifications contained the desired character (i.e., one row and one
column). The event-related potential (ERP) components evoked by these target
stimuli are different from those ERPs evoked by stimuli that did not contain the
desired character. The ERPs are composed of a combination of visual and
cognitive components. The subject’s task was to focus her/his attention on
characters (i.e. one at a time) in a word that was prescribed by the
investigator. For each character of the word, the 12 intensifications were
repeated 15 times before moving on to the next character. Any specific row or
column was intensified 15 times per character and there were in total 180
intensifications per character. The data was recorded using 64 channel EEG. The
64 channels covered the whole scalp of the subject and were aligned according to
the 10-20 system. The collected signals were bandpass filtered from 0.1-60Hz and
digitized at 240Hz. The data set consists of a training set of 85 characters and
a test set of 100 characters for each of the two subjects. For the trainings
sets the labels of the characters were available. The task for this data set was
to predict the labels of the test sets using the training sets and the labels.
In this experiment we reached a classification accuracy for single letters of
93,5\%, the winners of the competition reached 96,5\%, 90,5\%, and 90\%.

We also provide an example online experiment where we use the ERP data set with
an pseudo amplifier that feeds the data in real-time to the toolbox, to show how
to do the classification task in an online setting.

The data sets from the competition are freely available and one can reproduce
our results using the scripts and the data.

\section{Python 2 vs Python 3%
  \label{python-2-vs-python-3}%
}

The ongoing transition from Python 2 to Python 3 was also considered and we
decided to support \emph{both} Python versions. Wyrm is mainly developed under Python
2.7, but written in a \emph{forward compatible} way to support Python 3 as well. Our
unit tests ensure that the methods provide the expected results in Python 2 and
Python 3.

\section{Summary and Conclusion%
  \label{summary-and-conclusion}%
}

In this paper we presented Wyrm, a free and open source BCI toolbox in Python.
We introduced Wyrm's main data structure and explained the design ideas behind
the current implementation. We gave a short overview of the existing methods in
the toolbox and showed how we utilized unit testing to make sure the toolbox
works as specified, where to find the extensive documentation and some detailed
examples.

Together with \hyperref[mushu]{Mushu} \cite{Mushu} our signal acquisition library and Pyff \cite{Pyff}
our Framework for Feedback and Stimulus Presentation, Wyrm adds the final piece
to our ongoing effort to provide a complete, free and open source BCI system in
Python.

Wyrm is available under the terms of the MIT license, its repository can be
found at \url{http://github.com/venthur/wyrm}.

\section{Acknowledgements%
  \label{acknowledgements}%
}

This work was supported in part by grants of the BMBF: 01GQ0850 and 16SV5839.
The research leading to this results has received funding from the European
Union Seventh Framework Programme (FP7/2007-2013) under grant agreements 611570
and 609593.


\begin{thebibliography}{Scikit-learn}
\bibitem[Blankertz]{Blankertz}{

Blankertz B, Lemm S, Treder MS, Haufe S, Müller KR (2011)
\emph{Single-trial analysis and classification of ERP components – a
tutorial}. NeuroImage 56:814– 825,
\url{http://dx.doi.org/10.1016/j.neuroimage.2010.06.048}}
\bibitem[Dähne]{Dähne}{

Dähne S, Meinecke FC, Haufe S, Höhne J, Tangermann M, Müller KR,
Nikulin VV (2014) \emph{SPoC: a novel framework for relating the amplitude
of neuronal oscillations to behaviorally relevant parameters}.
NeuroImage 86(0):111–122,
\url{http:://dx.doi.org/10.1016/j.neuroimage.2013.07.079}}
\bibitem[Mushu]{Mushu}{

Bastian Venthur and Benjamin Blankertz. \emph{Mushu, a Free and Open
Source BCI Signal Acquisition, Written in Python}. Engineering in
Medicine and Biology Society (EMBC). doi:
\url{http://dx.doi.org/10.1109/EMBC.2012.6346296} San Diego, 2012.}
\bibitem[Pyff]{Pyff}{

Bastian Venthur, Simon Scholler, John Williamson, Sven Dähne, Matthias
S Treder, Maria T Kramarek, Klaus-Robert Müller and Benjamin
Blankertz. \emph{Pyff-{}-{}-A Pythonic Framework for Feedback Applications and
Stimulus Presentation in Neuroscience}. Frontiers in Neuroscience.
2010. \url{http://dx.doi.org/10.3389/fnins.2010.00179}.}
\bibitem[Ramoser]{Ramoser}{

Ramoser H, Muller-Gerking J, Pfurtscheller G (2000) \emph{Optimal
spatial filtering of single trial eeg during imagined hand
movement}. Rehabilitation Engineering, IEEE Transactions on
8(4):441–446}
\bibitem[Donchin]{Donchin}{

E. Donchin, K. Spencer, and R. Wijesinghe. \emph{The mental prosthesis:
assessing the speed of a p300-based brain-computer interface.}
Rehabilitation Engineering, IEEE Transactions on, 8(2):174–179, Jun
2000.}
\bibitem[Gramfort]{Gramfort}{

Gramfort, A., Luessi, M., Larson, E., Engemann, D. A., Strohmeier,
D., Brodbeck, C., ... \& Hämäläinen, M. (2013). \emph{MEG and EEG data
analysis with MNE-Python}. Frontiers in neuroscience, 7.}
\bibitem[Billinger]{Billinger}{

Billinger, M., Brunner, C., \& Müller-Putz, G. R. (2014). \emph{SCoT: a
Python toolbox for EEG source connectivity}. Frontiers in
neuroinformatics, 8. ISO 690}
\bibitem[Numpy]{Numpy}{

\url{http://numpy.org}}
\bibitem[SciPy]{SciPy}{

\url{http://scipy.org}}
\bibitem[Matplotlib]{Matplotlib}{

\url{http://matplotlib.org}}
\bibitem[Scikit-learn]{Scikit-learn}{

\url{http://scikit-learn.org}}
\bibitem[Pandas]{Pandas}{

\url{http://pandas.pydata.org}}
\bibitem[BCIComp3]{BCIComp3}{

\url{https://www.bbci.de/competition/iii/}}
\end{thebibliography}
\end{document}